\documentclass[aps,prl,reprint,superscriptaddress,floatfix,nofootinbib,longbibliography]{revtex4-2}

\usepackage{amsmath,amssymb}
\usepackage{amsthm}
\usepackage{graphicx}

\DeclareMathOperator{\spec}{spec}
\DeclareMathOperator{\dom}{Dom}
\newcommand{\R}{\mathbb{R}}
\newcommand{\C}{\mathbb{C}}
\newcommand{\Z}{\mathbb{Z}}
\newcommand{\HBBM}{H_{\mathrm{BBM}}}
\newcommand{\HBBMt}{\widetilde H_{\mathrm{BBM}}}
\newcommand{\Dt}{\overline D}
\newcommand{\Delt}{\overline\Delta}
\newcommand{\Dom}{\mathcal D_0}
\newcommand{\eps}{\varepsilon}

\begin{document}

\title{Metric completion of the Bender--Brody--M\"uller Hamiltonian:\\
dilation spectrum and missing eigenstates}

\author{Kejun Liu}
\email{kjliu@suda.edu.cn}
\affiliation{State Key Laboratory of Bioinspired Interface Material Science, Institute of Nano \& Functional Materials, Soochow University, Suzhou 215123, China}

\date{\today}

\begin{abstract}
The Bender--Brody--M\"uller (BBM) Hamiltonian was proposed as a non-Hermitian Hilbert--P\'olya operator.  We analyze the Hilbert completion induced, on the standard half-line $L^2$ core, by BBM's candidate metric $\hat\eta=\sin^2(\hat p/2)=\Delta^\dagger\Delta/4$.  The form $\eta_0=\Delta^\dagger\Delta$ is positive with trivial kernel but is not coercive.  Completing $C_c^\infty(0,\infty)$ in the norm $\|\psi\|_{\eta_0}=\|\Delta\psi\|$ gives a Hilbert space canonically unitarily equivalent to $L^2(\R_+)$; its free self-adjoint realization is the dilation generator, with simple, purely absolutely continuous spectrum $\R$.  The analysis yields two spectral statements of interest beyond the BBM problem.  First, no bounded sandwich $\Delta^\dagger h(D)\Delta$ is boundedly invertible.  Second, the transported symmetric operator has deficiency indices $(\infty,\infty)$ and an adjoint with every real point as an eigenvalue of infinite multiplicity, while its free extension is purely continuous.  The realization-independent BBM conclusion concerns the candidate eigenfunctions: $\Delta\psi_z=x^{-z}$, so for $\operatorname{Re}z=1/2$ they do not belong to the completed space.  Thus the original BBM boundary-condition/eigenfunction mechanism cannot produce point-spectrum Riemann-zero states in this $L^2$-based metric completion.
\end{abstract}

\maketitle

The Hilbert--P\'olya idea asks for a self-adjoint operator whose spectrum gives the imaginary parts of the nontrivial zeros of $\zeta(s)$.  It is one of the reasons why random-matrix statistics, $xp$ Hamiltonians, trace formulas, and de Branges spaces keep reappearing around the Riemann hypothesis~\cite{berry-keating-1999,connes1999,deBranges1986,conrey-li-1998,schumayer-hutchinson2011}.  The usual difficulty is not only to write a formal Hamiltonian with the right scaling behavior.  One must also identify a Hilbert space, a domain, a self-adjoint realization, and a mechanism that turns a continuum of scale transformations into a discrete sequence.

In 2017 Bender, Brody, and M\"uller (BBM) proposed a particularly sharp non-Hermitian candidate~\cite{bbm2017}.  In the notation used below, their formal Hamiltonian is
\begin{equation}
\label{eq:bbm}
\HBBM=(1-e^{-i\hat p})^{-1}(\hat x\hat p+\hat p\hat x)(1-e^{-i\hat p})
      =\Delta^{-1}D\Delta ,
\end{equation}
where
\begin{equation}
\label{eq:shift-d}
(\hat S\psi)(x)=\psi(x-1)\mathbf 1_{x>1},\qquad
\Delta=1-\hat S ,
\end{equation}
on $L^2(\R_+)$, and
\begin{equation}
D=\hat x\hat p+\hat p\hat x=-i(2x\partial_x+1)
\end{equation}
is the generator of dilations.  The proposed physical interpretation was that a suitable pseudo-Hermitian Hilbert space would make \eqref{eq:bbm} a real-spectrum Hamiltonian with eigenvalues proportional to the Riemann ordinates: for $z=\frac12+i\gamma$, the BBM normalization gives $E=i(2z-1)=-2\gamma$.  This is a natural physical move: in pseudo-Hermitian quantum mechanics a non-Hermitian Hamiltonian can represent ordinary unitary physics if there is a positive, invertible metric operator $\eta$ for which $H^\dagger\eta=\eta H$~\cite{mostafazadeh2002}.  The metric, however, plays a structural role: it fixes the topology of the physical state space.

The proposal immediately raised a precise set of structural questions.  Bellissard showed that $\hat p$ has no self-adjoint extension on the half-line, that the BBM eigenfunctions are not $L^2$ states on the critical line, and that a weighted-space cure destroys the critical line~\cite{bellissard2017}.  Moxley observed that a formal state redefinition conjugates the BBM eigenvalue equation to a dilation equation~\cite{moxley2017}.  BBM replied that the operator should be understood in pseudo-Hermitian quantum mechanics and biorthogonal quantum theory, where square-integrability is understood relative to the metric inner product rather than the standard one~\cite{bbm-response2017,brody-biorthogonal}.  Bender and Brody later left the momentum-space formulation of the eigenvalue problem as a challenging open problem~\cite{bender-brody2018}.  Recent Hilbert--P\'olya Hamiltonians use different operators and similarity structures and do not address the BBM shift metric~\cite{yakaboylu2024}.  None of these works constructed the standard $L^2$-based completion of BBM's candidate metric or determined the operators it supports.

BBM themselves proposed the candidate metric: in their notation $\hat\eta=\sin^2(\hat p/2)$, which on the half-line is exactly $\Delta^\dagger\Delta/4$.  They observed that it is positive and bounded but that its inverse is unbounded, and that their eigenfunctions are not normalizable in the standard inner product, which they argued does not affect the pseudo-Hermitian interpretation~\cite{bbm2017,bbm-response2017}.  We study the Hilbert completion generated by this bounded form from the standard dense core $C_c^\infty(0,\infty)\subset L^2(\R_+)$.  This is the natural $L^2$-based quasi-Hermitian construction~\cite{scholtz1992,mostafazadeh2010}; it does not purport to classify non-$L^2$, rigged-Hilbert-space, or distributional formulations.  A positive metric with unbounded inverse need not define an equivalent Hilbert topology, and its completion may change which candidate states exist, as in exactly solvable non-Hermitian models with the same obstruction~\cite{siegl-krejcirik2012}.  The analysis below uses only BBM's similarity structure, their candidate metric, and this standard completion procedure.

For the free transported realization the result is the dilation continuum.  More generally, the completion admits many self-adjoint realizations, but none can contain the BBM candidate eigenfunctions as Hilbert-space vectors on the critical line.

The result has two layers.  The BBM-specific layer determines the completed state space and its operator realizations: the completion is canonically $L^2(\R_+)$ in the coordinate $\Phi=\Delta\psi$, the free transported realization is the self-adjoint dilation generator, and the original BBM eigenfunctions are absent because their shift differences are the non-$L^2$ power functions $x^{-z}$ on the critical line.  Two further statements hold independently of the zeta-function proposal.  First, the loss of coercivity is universal within the bounded-sandwich class: no $\Delta^\dagger h(D)\Delta$ with bounded $h$ is boundedly invertible.  Second, the transported minimal operator has deficiency indices $(\infty,\infty)$; its free extension is purely continuous, whereas its adjoint has every real point as an eigenvalue of infinite multiplicity.  The proof below follows this order.

\begin{figure}[t]
\includegraphics[width=\columnwidth]{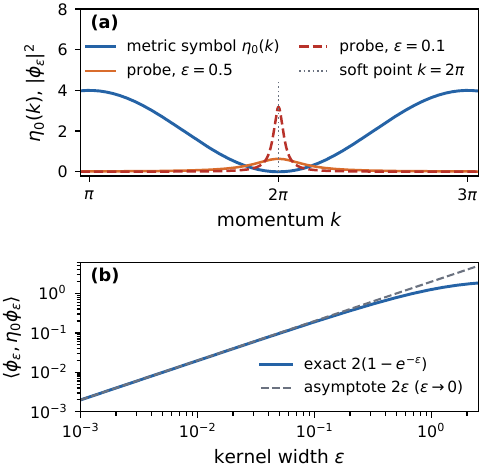}
\caption{\label{fig:metric-collapse}
Metric degeneracy of the BBM form $\eta_0=\Delta^\dagger\Delta$ in the Hardy representation.  (a) Local window $k\in[\pi,3\pi]$ around the soft point $k=2\pi$: the boundary symbol $\eta_0(k)=2(1-\cos k)$ vanishes quadratically at $k=2\pi\mathbb Z$, while the normalized Cauchy-kernel probes $|\phi_\varepsilon|^2$ concentrate at the soft point as $\varepsilon$ decreases.  (b) The squared $\eta_0$-norm of the unit-$L^2$-norm probe, $\langle\phi_\varepsilon,\eta_0\phi_\varepsilon\rangle=2(1-e^{-\varepsilon})$ (exact residue evaluation), vanishes linearly in the small-width limit, $2(1-e^{-\varepsilon})=2\varepsilon+O(\varepsilon^2)$ as $\varepsilon\to0$: normalized states can carry arbitrarily small metric length, i.e. $\eta_0$ is not coercive.}
\end{figure}

\paragraph{The natural metric.}
The similarity form \eqref{eq:bbm} canonically selects the positive form
\begin{equation}
\label{eq:eta0}
\eta_0=\Delta^\dagger\Delta=2I-\hat S-\hat S^\dagger .
\end{equation}
This is the precise half-line meaning of the physics shorthand $2(1-\cos\hat p)$.  On the test domain
\begin{equation}
\Dom=C_c^\infty(0,\infty)
\end{equation}
the shift preserves $\Dom$.  Indeed, if $\psi$ is smooth and compactly supported away from the origin, then $\psi(x-1)\mathbf 1_{x>1}$ is again smooth and compactly supported away from the origin.  Hence $\Delta\psi\in\Dom$.  The similarity notation suggests
\begin{equation}
\HBBM^\dagger\eta_0=\eta_0\HBBM\qquad\text{(formal)}.
\end{equation}
Before an operator domain is chosen, neither side is an operator identity.  Its rigorous content on the test space is the symmetry of $D$ on $\Delta\Dom$; the resulting densely defined operator is $T=\Dt|_{\Delta\Dom}$ below.  The first question is whether the positive form defines a Hilbert topology equivalent to the ambient one.  In the usual bounded-metric setting it must be coercive.  The BBM form fails precisely there.

Extend functions on $\R_+$ by zero to $\R$.  With the Fourier convention
\[
(\mathcal F\psi)(k)=\frac{1}{\sqrt{2\pi}}\int_\R e^{-ikx}\psi(x)\,dx,
\]
Paley--Wiener gives the unitary identification
\begin{equation}
\label{eq:pw-main}
\mathcal F:L^2(\R_+)\simeq H^2(\C^-).
\end{equation}
In this representation $\Delta$ is boundary multiplication by $1-e^{-ik}$.  The quadratic form of \eqref{eq:eta0} has boundary symbol
\begin{equation}
\label{eq:eta-symbol}
\eta_0(k)=|1-e^{-ik}|^2=2(1-\cos k).
\end{equation}
That is, $\langle\psi,\eta_0\psi\rangle=\int_\R\eta_0(k)|\mathcal F\psi(k)|^2\,dk$.  Its zeros at $2\pi\Z$ do not give $L^2$ null vectors, but they do give soft directions of the metric.  This is the first indication that the metric topology is weaker than the ambient one.

Physically, $\Delta$ differentiates against a unit shift, so slowly varying wave packets---Fourier-concentrated near the zeros of $1-e^{-ik}$---have arbitrarily small metric length.  The metric therefore changes the topology; its spectral realizations must be analyzed separately.

\paragraph{Metric degeneracy.}
For $\eps>0$ set
\begin{equation}
\label{eq:cauchy}
\phi_\eps(k)=\frac{\sqrt{\eps/\pi}}{k-2\pi-i\eps}.
\end{equation}
The pole lies in $\C^+$, hence $\phi_\eps\in H^2(\C^-)$, and $\|\phi_\eps\|_{L^2(\R)}=1$.  A residue calculation gives
\begin{align}
\label{eq:collapse-form}
\langle \phi_\eps,\eta_0\phi_\eps\rangle
&=\frac{2\eps}{\pi}\int_{\R}\frac{1-\cos u}{u^2+\eps^2}\,du \nonumber\\
&=2(1-e^{-\eps})\longrightarrow0 .
\end{align}
Since the Fourier images of $C_c^\infty(0,\infty)$ are dense in $H^2(\C^-)$ and the form is bounded above by $4$, the same zero infimum is obtained on the original BBM test domain:
\begin{equation}
\label{eq:inf0}
\inf_{\psi\in \Dom,\,\|\psi\|=1}\langle\psi,\eta_0\psi\rangle=0 .
\end{equation}
Thus the BBM metric is positive and has trivial kernel, but it is not bounded below.  This is stronger than saying that a formal differential expression has a domain problem.  The physical metric topology selected by the BBM similarity transform is itself noncoercive.  The symbol and the degeneracy of the form are shown in Fig.~\ref{fig:metric-collapse}.

The Cauchy kernel stays inside the half-line Hardy space: the probe is a legitimate Paley--Wiener state, normalized in $L^2$, whose metric length goes to zero; no full-line plane waves, which lie outside the Hardy space, are used.

It is useful to separate this from an eigenvector statement.  The zero set $2\pi\Z$ has measure zero, so the multiplier on the ambient $L^2(\R,dk)$ has no nonzero kernel.  The degeneracy is form-level: the form assigns arbitrarily small metric length to normalized states.

\paragraph{A universal obstruction.}
One may try to insert an additional observable in the dilation representation.  Let $h$ be bounded and measurable and define the metric sandwich
\begin{equation}
\label{eq:bounded-sandwich}
\eta_h=\Delta^\dagger h(D)\Delta .
\end{equation}
This includes the positive case $h\ge 0$ and the natural indefinite attempt $h=\mathrm{sgn}$ in the Mellin spectral variable.  The repair never produces a boundedly invertible metric.

The physical obstruction is a symmetry mismatch.  The degeneracy of $\Delta^\dagger\Delta$ is localized in momentum: the symbol $|1-e^{-ik}|^2$ vanishes at $k\in2\pi\Z$.  A multiplier $h(D)$, however, belongs to the spectral algebra of scale transformations.  Momentum and scale cannot be diagonalized together.  With our normalization of $D=xp+px$ one has
\begin{equation}
[D,p]=2ip ,
\end{equation}
or $[D/2,p]=ip$ for the half-dilation generator.  Thus a bounded function of $D$ cannot cancel the momentum zeros of $\Delta$ pointwise: in the momentum representation it acts across scales rather than near the set $2\pi\Z$.  Cancelling them pointwise would require an unbounded operation in the wrong spectral variable---not a bounded metric deformation.

Indeed choose $\chi\in C_c^\infty(1,2)$ with $\|\chi\|=1$ and set
\begin{equation}
\psi_n(x)=n^{-1/2}\chi(x/n).
\end{equation}
Then $\|\psi_n\|=1$, while translation continuity gives
\begin{equation}
\label{eq:wide-state}
\|\Delta\psi_n\|_{L^2(\R_+)}
 =\|\psi_n-\hat S\psi_n\|\longrightarrow0 .
\end{equation}
Since $h(D)$ and $\Delta$ are bounded,
\begin{equation}
\label{eq:no-repair}
\|\eta_h\psi_n\|
\le \|\Delta^\dagger\|\,\|h(D)\|\,\|\Delta\psi_n\|
\longrightarrow0 .
\end{equation}
Hence $\eta_h$ is not bounded below on the unit sphere and cannot have a bounded inverse.  In particular, the indefinite form $\Delta^\dagger \mathrm{sgn}(D)\Delta$ is not a classical Krein or Pontryagin metric with a bounded fundamental symmetry~\cite{bognar1974}.  This also lies outside the usual bounded-invertible metric setting of pseudo-Hermitian quantum mechanics~\cite{mostafazadeh2002}.

This obstruction is universal: it is independent of the choice of $h$.  Even for $h$ bounded away from zero, the outer factors $\Delta^\dagger$ and $\Delta$ still make broad slowly varying states almost invisible.  The obstruction is therefore not caused by the sign, phase, or positivity of the inserted observable.  It is caused by the shift-difference factor that BBM needs in order to connect the Hurwitz recurrence to the dilation generator.

For the indefinite choice $h=\mathrm{sgn}$, one can say more.  The operator
\begin{equation}
\eta_{\mathrm{sgn}}=\Delta^\dagger\mathrm{sgn}(D)\Delta
\end{equation}
is self-adjoint and indefinite, because $\Delta\Dom$ is dense and $\mathrm{sgn}(D)$ has nonzero positive and negative spectral subspaces.  The same broad states give $0\in\spec(\eta_{\mathrm{sgn}})$.  But $0$ is not an eigenvalue: if $\eta_{\mathrm{sgn}}\psi=0$, density gives $\ker\Delta^\dagger=\{0\}$, hence $\mathrm{sgn}(D)\Delta\psi=0$, and since $\mathrm{sgn}(D)^2=I$, $\Delta\psi=0$.  The equation $\Delta\psi=0$ says $\psi(x)=\psi(x-1)$ for $x>1$ and $\psi=0$ on $(0,1)$, so $\psi=0$.  Thus $0$ lies in the continuous spectrum of the would-be Krein metric.  This is exactly the regime where the usual indefinite-metric spectral arguments lose their main hypothesis.

\paragraph{The $L^2$-based completion.}
We now complete the standard test space in the norm defined by \eqref{eq:eta0}:
\begin{equation}
\label{eq:eta-norm}
\|\psi\|_{\eta_0}=\langle\psi,\eta_0\psi\rangle^{1/2}
                 =\|\Delta\psi\|_{L^2(\R_+)},
        \qquad \psi\in\Dom .
\end{equation}
Let $\widetilde{\mathcal H}_{\eta_0}$ be this completion.  The map $\Delta:\Dom\to L^2(\R_+)$ is an isometry from the $\eta_0$ norm to the ordinary $L^2$ norm.  Its image is dense.  To see this, if $g\perp\Delta\Dom$, then $\Delta^\dagger g=0$, i.e.
\begin{equation}
g(x)=g(x+1)\quad \hbox{a.e. on }\R_+ .
\end{equation}
The only $L^2(\R_+)$ function with this one-periodicity is zero, since otherwise
\begin{align}
\int_0^\infty |g(x)|^2\,dx
&=\sum_{n=0}^\infty\int_0^1 |g(x+n)|^2\,dx \nonumber\\
&=\sum_{n=0}^\infty\int_0^1 |g(x)|^2\,dx
\end{align}
would diverge.  Thus $\Delta\Dom$ is dense, and the isometry extends to a unitary map
\begin{equation}
\label{eq:unitary-delta}
\Delt:\widetilde{\mathcal H}_{\eta_0}\longrightarrow L^2(\R_+).
\end{equation}
Thus this $\eta_0$ completion is canonically $L^2(\R_+)$, with coordinate map given by the shift difference $\Delta$.  This determines the completed state space, but not a self-adjoint realization of the formal Hamiltonian.

In physical language, the completed theory has already chosen its observable wavefunction:
\begin{equation}
\Phi=\Delt\psi=\Delta\psi .
\end{equation}
Once the state is written as $\Phi$, the similarity reduces the formal action to that of the dilation expression.  It still leaves the operator domain open.  The free choice below makes rigorous the reduction to the Berry--Keating Hamiltonian that BBM noted when conjugating with $\rho=\Delta$~\cite{bbm2017}.

Let $\Dt$ be the self-adjoint closure of $D$ on $C_c^\infty(0,\infty)$.  Choose the free transported realization
\begin{equation}
\label{eq:completed-h}
\HBBMt=\Delt^{-1}\Dt\Delt,\qquad
\dom(\HBBMt)=\Delt^{-1}\dom(\Dt).
\end{equation}
For every $\psi\in\Dom$ one has $\Delta\psi\in\Dom$, hence
\begin{equation}
\label{eq:compat}
\Delt\,\HBBMt\psi=D\Delta\psi .
\end{equation}
Thus \eqref{eq:completed-h} is one rigorous realization of the formal expression $\Delta^{-1}D\Delta$ in the metric completion, without requiring $\Delta^{-1}$ to be a bounded operator on all of $L^2$.

For this choice, the formal cancellation becomes rigorous.  On the original $L^2$ space $\Delta^{-1}$ is not bounded and everywhere defined; between the completed space and $L^2$, however, $\Delt^{-1}$ is unitary.  Equation~\eqref{eq:completed-h} therefore has the spectrum of $\Dt$.

\paragraph{Spectrum of the free realization.}
Under the Mellin transform,
\begin{equation}
\label{eq:mellin}
(\mathcal M D\mathcal M^{-1}f)(\tau)=2\tau f(\tau).
\end{equation}
Therefore $\Dt$ has purely absolutely continuous spectrum $\R$ with multiplicity one.  Since \eqref{eq:completed-h} is a unitary equivalence,
\begin{equation}
\label{eq:main}
\spec(\HBBMt)=\R,\qquad
\spec_{\mathrm{pp}}(\HBBMt)=\varnothing ,
\end{equation}
and the spectrum is purely absolutely continuous with multiplicity one.  This realization has no isolated levels.

\paragraph{The adjoint's real point spectrum.}
The same construction exhibits a spectral phenomenon of interest beyond the BBM problem.  Let $T=\Dt|_{\Delta\Dom}$.  Under the unitary \eqref{eq:unitary-delta}, this is the symmetric operator generated by the formal BBM rule on $\Dom$; its self-adjoint extensions are the corresponding realizations in the completed space.  The free extension is $\Dt$.  The Supplemental Material proves that $T$ has deficiency indices $(\infty,\infty)$ and that $\ker(T^*-\lambda)$ is infinite-dimensional for every $\lambda\in\R$.  Thus the same symmetric core has a purely continuous free extension while its adjoint has the whole real line as point spectrum.  In particular, the metric completion does not select the free extension, and other extension data may introduce point spectrum, as in related $xp$ models~\cite{sierra-rl2011,schumayer-hutchinson2011}.

\paragraph{The missing eigenstates.}
What decides the Hilbert--P\'olya question is not which realization one prefers, but whether the BBM eigenstates live in the completed space at all.  They do not.  For $\psi_z(x)=-\zeta(z,x+1)$ the Hurwitz identity $\zeta(z,x)-\zeta(z,x+1)=x^{-z}$ gives
\begin{equation}
\label{eq:hurwitz}
(\Delta\psi_z)(x)=x^{-z},\qquad x>1 .
\end{equation}
On the critical line $|x^{-z}|=x^{-1/2}$, so $\Delta\psi_z\notin L^2(\R_+)$: the candidate eigenfunction has no representative in $\widetilde{\mathcal H}_{\eta_0}$, whose elements are exactly the states with $L^2$ shift-difference.  This rules out the original BBM mechanism---the specified functions $\psi_z$ together with the boundary condition $\psi_z(0)=0$---for every self-adjoint realization in this completed space, not only for the free realization.  No such realization can use the BBM functions as Riemann-zero eigenvectors.  Other extensions may have unrelated point spectrum, but they cannot restore these missing vectors.

\paragraph{Remark on de Branges-space realizations.}
The bounded $L^2$-based analysis neither excludes nor constructs an unbounded-metric, rigged-Hilbert-space, or de Branges realization.  For the free transported choice, a separate de Branges reformulation would face the classical de Branges problem~\cite{deBranges1968,deBranges1986,conrey-li-1998}.  The immediate candidate $E(z)=\xi(\frac12+iz)$ satisfies $E^\#=E$, so it fails the strict Hermite--Biehler inequality; this observation does not classify other choices of $E$ (Supplemental Material).

This locates the new content relative to the earlier comments~\cite{bellissard2017,moxley2017,bbm-response2017,bender-brody2018}, which identified the similarity to the dilation generator and raised the associated domain questions.  What the present analysis adds is the completion generated by BBM's metric, the nonuniqueness of the operator realization on that completion, and the failure of the original BBM candidate states to survive it.

The physics message is that the metric fixes a topology, not a Hamiltonian domain.  BBM's metric is noncoercive, and its standard $L^2$-based completion is unitarily $L^2(\R_+)$ in the coordinate $\Delta\psi$.  Choosing the free dilation realization gives a simple absolutely continuous spectrum, whereas the transported symmetric core has infinitely many self-adjoint extensions.  That freedom does not rescue the original BBM mechanism: for every extension in this completion, the proposed $\psi_z$ are absent and the boundary condition on them cannot generate Riemann-zero eigenstates.  The theorem therefore rules out the original BBM candidate-eigenfunction/boundary-condition mechanism, not every construction inspired by it.  A surviving route must leave at least one hypothesis of the bounded $L^2$-based construction studied here, for example by using a non-$L^2$ or rigged Hilbert space, an unbounded metric, a distributional state space, or a genuinely new operator domain or eigenfunction ansatz.  Such a route would be a new construction, not a rescue by choosing another self-adjoint extension: it must specify its topology, state space, and operator domain and prove self-adjointness and the claimed spectrum independently.  The same quasi-invertibility issue occurs in exactly solvable non-Hermitian models~\cite{siegl-krejcirik2012}; here it is forced by the BBM shift factor, and no bounded insertion $h(D)$ repairs it.  Complete proofs are given in the Supplemental Material.

\begin{acknowledgments}
This work was supported by the National High-Level Overseas Talent Program (KS21400126), the Suzhou Talent project (ZXP2025057), the Jiangsu Distinguished Professorship Fund (SR21400225), and the Research Start-up Fund (NH21400525).
\end{acknowledgments}

\paragraph{Data availability.}
No data were generated or analyzed in this theoretical study.

\end{document}

\title{Supplemental Material:\\
Metric completion of the Bender--Brody--M\"uller Hamiltonian:\\
dilation spectrum and missing eigenstates}

\author{Kejun Liu}
\email{kjliu@suda.edu.cn}
\affiliation{State Key Laboratory of Bioinspired Interface Material Science, Institute of Nano \& Functional Materials, Soochow University, Suzhou 215123, China}

\date{\today}
\maketitle

\noindent This Supplemental Material gives the operator-domain details behind the Letter
for the standard half-line $L^2$ core and its metric completion.
Equation, figure, and table numbers carry an ``S'' prefix; references to
unprefixed items point to the main text.

\setcounter{equation}{0}
\setcounter{theorem}{0}
\renewcommand{\theequation}{S\arabic{equation}}
\renewcommand{\thetheorem}{S\arabic{theorem}}

We use
\[
\R_+=(0,\infty),\qquad \Dom=C_c^\infty(0,\infty)\subset L^2(\R_+),
\]
and define
\[
(\hat S\psi)(x)=\psi(x-1)\mathbf 1_{x>1},\qquad
\Delta=1-\hat S,\qquad
D=-i(2x\partial_x+1).
\]
The formal BBM expression is $\HBBM=\Delta^{-1}D\Delta$.
Our Hilbert-space inner product is
$\langle u,v\rangle=\int_0^\infty\overline{u(x)}v(x)\,dx$, linear in the second argument.

\section{Half-line shift algebra}

\begin{lemma}[$\Delta$ preserves the test domain]
\label{lem:S-delta-domain}
For every $\psi\in\Dom$, one has $\Delta\psi\in\Dom$.
\end{lemma}

\begin{proof}
The support of $\psi\in C_c^\infty(0,\infty)$ has positive distance from the origin.  The shifted function
\[
(\hat S\psi)(x)=\psi(x-1)\mathbf 1_{x>1}
\]
is therefore again smooth and compactly supported in $(0,\infty)$, with no jump at $x=1$.  Hence $\Delta\psi=\psi-\hat S\psi\in\Dom$.
\end{proof}

The natural pseudo-Hermitian form selected by the similarity transform is
\begin{equation}
\label{eq:S-eta0}
\eta_0=\Delta^\dagger\Delta=2I-\hat S-\hat S^\dagger .
\end{equation}
It is bounded and positive on $L^2(\R_+)$.  The similarity notation gives the mnemonic
\[
\HBBM^\dagger\eta_0=\eta_0\HBBM\qquad\text{(formal)},
\]
because $\Delta\HBBM=D\Delta$ and $D$ is symmetric on $\Delta\Dom\subset\Dom$.
At this stage no domain has been assigned to $\HBBM$ or its adjoint, so this is not
an operator identity.  The rigorous symmetric operator carrying the same rule is
$T=\Dt|_{\Delta\Dom}$ in Eq.~\eqref{eq:S-T} below.

Extend functions on $\R_+$ by zero and take
\[
(\mathcal F\psi)(k)=\frac{1}{\sqrt{2\pi}}\int_\R e^{-ikx}\psi(x)\,dx .
\]
Paley--Wiener gives the unitary map
\begin{equation}
\label{eq:S-PW}
\mathcal F:L^2(\R_+)\longrightarrow H^2(\C^-).
\end{equation}
Under this map $\Delta$ becomes boundary multiplication by $1-e^{-ik}$.  Consequently
the quadratic form $q_0[\psi]=\langle\psi,\eta_0\psi\rangle$ has boundary representation
\[
q_0[\psi]=\int_\R \eta_0(k)|\mathcal F\psi(k)|^2\,dk,
\]
with symbol
\begin{equation}
\label{eq:S-eta-symbol}
\eta_0(k)=|1-e^{-ik}|^2=2(1-\cos k).
\end{equation}

\section{Form degeneracy of the natural metric}

\begin{theorem}[Form degeneracy]
\label{thm:S-form-degeneracy}
For the boundary-symbol form $q_0$ above,
\begin{equation}
\label{eq:S-inf-zero}
\inf_{\psi\in H^2(\C^-),\,\|\psi\|=1}
\langle \psi,\eta_0\psi\rangle=0 .
\end{equation}
The same infimum is obtained on the Fourier image of $\Dom$.
\end{theorem}

\begin{proof}
For $\eps>0$ define
\begin{equation}
\label{eq:S-cauchy}
\phi_\eps(k)=\frac{\sqrt{\eps/\pi}}{k-2\pi-i\eps}.
\end{equation}
Its pole lies in $\C^+$, so $\phi_\eps\in H^2(\C^-)$, and
\[
\|\phi_\eps\|_{L^2(\R)}^2=
\frac{\eps}{\pi}\int_\R\frac{du}{u^2+\eps^2}=1 .
\]
Using $u=k-2\pi$,
\begin{align}
\langle\phi_\eps,\eta_0\phi_\eps\rangle
&=\frac{2\eps}{\pi}\int_\R
\frac{1-\cos u}{u^2+\eps^2}\,du \nonumber\\
&=2(1-e^{-\eps})\to0.
\end{align}
The last equality follows from the standard residue evaluation
\[
\int_\R\frac{1-\cos u}{u^2+\eps^2}\,du
=\frac{\pi}{\eps}(1-e^{-\eps}).
\]
Since $\mathcal F\Dom$ is dense in $H^2(\C^-)$ and the multiplier is bounded by $4$, the same infimum holds on $\mathcal F\Dom$.
\end{proof}

\begin{remark}
As a multiplication operator on $L^2(\R,dk)$, $\eta_0(k)$ has trivial kernel because $2\pi\Z$ has measure zero.  The degeneracy above is therefore not a null-vector statement.  It is a form-topology statement: the positive form has soft directions and is not bounded below.
\end{remark}

\section{No bounded multiplier repairs the metric}

\begin{theorem}[Bounded-sandwich no-go]
\label{thm:S-sandwich}
Let $h:\R\to\C$ be bounded and measurable.  Then
\begin{equation}
\label{eq:S-sandwich}
\eta_h=\Delta^\dagger h(D)\Delta
\end{equation}
extends to a bounded operator on $L^2(\R_+)$, but it has no bounded inverse.
\end{theorem}

\begin{proof}
Boundedness follows from the boundedness of $\Delta$ and the functional calculus for the self-adjoint closure of $D$:
\[
\|\eta_h\|\le \|\Delta\|^2\|h\|_\infty .
\]
Choose $\chi\in C_c^\infty(1,2)$ with $\|\chi\|=1$ and set
\[
\psi_n(x)=n^{-1/2}\chi(x/n).
\]
Then $\|\psi_n\|=1$ and, for large $n$, translation continuity gives
\[
\|\Delta\psi_n\|=\|\psi_n-\hat S\psi_n\|\to0 .
\]
Therefore
\[
\|\eta_h\psi_n\|\le
\|\Delta^\dagger\|\,\|h(D)\|\,\|\Delta\psi_n\|\to0 .
\]
Thus $\eta_h$ is not bounded below on the unit sphere and cannot be boundedly invertible.
\end{proof}

\begin{corollary}[Indefinite metric]
\label{cor:S-indefinite}
The operator $\eta_{\mathrm{sgn}}=\Delta^\dagger\mathrm{sgn}(D)\Delta$ is bounded, self-adjoint, and indefinite, but $0\in\spec_{\mathrm{cont}}(\eta_{\mathrm{sgn}})$.
\end{corollary}

\begin{proof}
It is indefinite because $\Delta\Dom$ is dense in $L^2(\R_+)$, while $\mathrm{sgn}(D)$ has nonzero positive and negative spectral subspaces.  The sequence in the proof of Theorem~\ref{thm:S-sandwich} gives $0\in\spec(\eta_{\mathrm{sgn}})$.  It remains to exclude a zero eigenvalue.  If $\eta_{\mathrm{sgn}}\psi=0$, then
\[
\Delta^\dagger\mathrm{sgn}(D)\Delta\psi=0 .
\]
Lemma~\ref{lem:S-density} below gives $\ker\Delta^\dagger=\{0\}$; hence $\mathrm{sgn}(D)\Delta\psi=0$.  Since $\mathrm{sgn}(D)^2=I$, $\Delta\psi=0$.  Then $\psi=\hat S\psi$, which forces $\psi=0$ on $(0,1)$ and recursively on all intervals $(n,n+1)$.  Hence $0$ is not an eigenvalue.  For a self-adjoint operator this places $0$ in the continuous spectrum.
\end{proof}

\section{The metric completion on the standard core}

Let $\widetilde{\mathcal H}_{\eta_0}$ be the completion of $\Dom$ in the norm
\begin{equation}
\label{eq:S-eta-norm}
\|\psi\|_{\eta_0}=\|\Delta\psi\|_{L^2(\R_+)} .
\end{equation}

\begin{lemma}[Density of the image]
\label{lem:S-density}
$\Delta\Dom$ is dense in $L^2(\R_+)$.
\end{lemma}

\begin{proof}
Suppose $g\perp\Delta\Dom$.  Then $\Delta^\dagger g=0$, i.e.
\[
g(x)=g(x+1)\quad\hbox{a.e. on }\R_+ .
\]
If $g$ is nonzero on a set of positive measure in $(0,1)$, the periodic repetition gives infinite $L^2$ norm on $\R_+$.  Hence $g=0$.  Thus $\ker\Delta^\dagger=\{0\}$ and $\overline{\ran\Delta}=L^2(\R_+)$.  Since $\Dom$ is dense and $\Delta$ is bounded, $\Delta\Dom$ is dense.
\end{proof}

\begin{theorem}[Completion isomorphism]
\label{thm:S-completion}
The map $\Delta:\Dom\to L^2(\R_+)$ extends to a unitary map
\begin{equation}
\Delt:\widetilde{\mathcal H}_{\eta_0}\longrightarrow L^2(\R_+).
\end{equation}
\end{theorem}

\begin{proof}
Equation \eqref{eq:S-eta-norm} says that $\Delta$ is isometric from $(\Dom,\|\cdot\|_{\eta_0})$ into $L^2(\R_+)$.  Lemma~\ref{lem:S-density} gives dense range.  The completion of an isometry with dense range is unitary.
\end{proof}

Let $\Dt$ be the self-adjoint closure of $D$ on $C_c^\infty(0,\infty)$ and choose the
free transported realization
\begin{equation}
\label{eq:S-completed-H}
\HBBMt=\Delt^{-1}\Dt\Delt,\qquad
\dom(\HBBMt)=\Delt^{-1}\dom(\Dt).
\end{equation}

\begin{proposition}[Compatibility with the formal BBM expression]
\label{prop:S-compat}
For every $\psi\in\Dom$,
\begin{equation}
\Delt\,\HBBMt\psi=D\Delta\psi .
\end{equation}
\end{proposition}

\begin{proof}
By Lemma~\ref{lem:S-delta-domain}, $\Delta\psi\in\Dom\subset\dom(\Dt)$.  Hence
\[
\Delt\,\HBBMt\psi=\Dt\,\Delt\psi=\Dt\,\Delta\psi=D\Delta\psi .
\]
Thus \eqref{eq:S-completed-H} is one rigorous metric-completion realization of $\Delta^{-1}D\Delta$.
\end{proof}

\begin{theorem}[Free spectral reduction]
\label{thm:S-main}
$\HBBMt$ is unitarily equivalent to $\Dt$.  Consequently
\begin{equation}
\spec(\HBBMt)=\R,\qquad
\spec_{\mathrm{pp}}(\HBBMt)=\varnothing ,
\end{equation}
and the spectrum is purely absolutely continuous with multiplicity one.
\end{theorem}

\begin{proof}
The unitary equivalence is \eqref{eq:S-completed-H}.  Under the Mellin transform, $D$ is multiplication by $2\tau$:
\[
(\mathcal M D\mathcal M^{-1}f)(\tau)=2\tau f(\tau).
\]
Thus $\Dt$ has purely absolutely continuous spectrum $\R$ with multiplicity one, and no point spectrum.  The same holds for $\HBBMt$.
\end{proof}

\section{Non-uniqueness: deficiency indices of the formal operator}
\label{sec:S-deficiency}

The spectral-reduction theorem describes the free realization \eqref{eq:S-completed-H}.  We now show that the formal operator itself is far from essentially self-adjoint, so that the free realization is a choice and not a consequence.  Define
\begin{equation}
\label{eq:S-T}
T=\Dt|_{\Delta\Dom},\qquad \dom(T)=\Delta\Dom .
\end{equation}
By Lemma~\ref{lem:S-density} its domain is dense, and $T\subset\Dt$ is symmetric.  Under the unitary $\Delt$ of Theorem~\ref{thm:S-completion}, $T$ is the image of the formal BBM expression on $\Dom$; its self-adjoint extensions are exactly the self-adjoint realizations of the formal expression in $\widetilde{\mathcal H}_{\eta_0}$.

For $g\in L^2_{\mathrm{loc}}(\R_+)$ and $\mu\in\C$, we use the weak distribution
$[D-\mu]g$ defined by
\begin{equation}
\label{eq:S-weak-pairing}
\langle [D-\mu]g,f\rangle_{\mathrm w}
:=\langle g,(D-\bar\mu)f\rangle,
\qquad f\in\Dom.
\end{equation}
For smooth $g$ this is the weak form of the usual function $(D-\mu)g$.

\begin{lemma}[Annihilators of the test domain]
\label{lem:S-annihilator}
A distribution $R$ on $(0,\infty)$ satisfies $\langle R,\Delta\psi\rangle=0$ for every $\psi\in\Dom$ if and only if $R$ is period-$1$: $R(x)=R(x+1)$ for $x>0$.
\end{lemma}

\begin{proof}
By the definition of a translated distribution,
$\langle R,\hat S\psi\rangle=\langle R(\cdot+1),\psi\rangle$.
Thus $\langle R,\Delta\psi\rangle=0$ for every test function if and only if
$R-R(\cdot+1)=0$ on $(0,\infty)$.
\end{proof}

\begin{lemma}[Deficiency criterion]
\label{lem:S-defcriterion}
For $\mu\in\C$, $g\in\ker(T^*-\mu)$ if and only if $g\in L^2(\R_+)$ and the distribution $(D-\mu)g$ is period-$1$.
\end{lemma}

\begin{proof}
With the inner-product convention stated above, $T^*g=\mu g$ is equivalent to
$\langle g,(D-\bar\mu)f\rangle=0$ for all $f\in\Delta\Dom$.  By
\eqref{eq:S-weak-pairing}, this says that $[D-\mu]g$ annihilates $\Delta\Dom$.
Lemma~\ref{lem:S-annihilator} is therefore exactly the required period-$1$ condition.
Conversely, that condition gives
$\langle g,Df\rangle=\bar\mu\langle g,f\rangle$ for every $f\in\Delta\Dom$,
which is the adjoint identity for $T^*g=\mu g$.
\end{proof}

\begin{theorem}[Deficiency indices]
\label{thm:S-deficiency}
$T$ has deficiency indices $(\infty,\infty)$.
\end{theorem}

\begin{proof}
For $\mu=i$ and $n\in\mathbb Z\setminus\{0\}$ set
\[
g_n(x)=\frac{1-e^{2\pi i n x}}{x} .
\]
Then $g_n(x)=O(1)$ at $0$ and $|g_n(x)|\le 2/x$ at $\infty$, so $g_n\in L^2(\R_+)$.  A direct computation gives
\[
(D-i)g_n=-2ixg_n'-2ig_n=-4\pi n\,e^{2\pi i n x},
\]
since $xg_n'=-2\pi i n e^{2\pi i n x}-(1-e^{2\pi i n x})/x$ and the non-periodic tail cancels between the two terms.  The right side is period-$1$, so $g_n\in\ker(T^*-i)$ by Lemma~\ref{lem:S-defcriterion}.

For $\mu=-i$ set
\[
\tilde g_n(x)=-\int_x^\infty \frac{e^{2\pi i n t}}{t}\,dt ,
\]
which is conditionally convergent.  Integration by parts gives $\tilde g_n(x)=e^{2\pi i n x}/(2\pi i n x)+O(x^{-2})$ at $\infty$ and $\tilde g_n(x)=\ln x+O(1)$ at $0$, so $\tilde g_n\in L^2(\R_+)$.  Moreover $x\tilde g_n'=e^{2\pi i n x}$, hence
\[
(D+i)\tilde g_n=-2ix\tilde g_n'=-2i\,e^{2\pi i n x},
\]
period-$1$, so $\tilde g_n\in\ker(T^*+i)$.  The first family is linearly independent:
after multiplying a finite relation by $x$, independence of the distinct Fourier modes
$1,e^{2\pi i n x}$ forces every coefficient to vanish.  For the second family,
differentiating a finite relation gives independence of the modes $e^{2\pi i n x}$.
Thus both deficiency spaces are infinite-dimensional, giving indices $(\infty,\infty)$.
\end{proof}

\begin{corollary}
\label{cor:S-nonunique}
The formal BBM expression has infinitely many self-adjoint realizations in $\widetilde{\mathcal H}_{\eta_0}$; the free realization \eqref{eq:S-completed-H} is one of them and is not selected by the formal expression alone.
\end{corollary}

\section{The adjoint has every real point as an eigenvalue}
\label{sec:S-realspec}

\begin{theorem}
\label{thm:S-realspec}
For every $\lambda\in\R$, $\ker(T^*-\lambda)$ is infinite-dimensional.
\end{theorem}

\begin{proof}
Set $s=(1+i\lambda)/2$ and
\[
R_\lambda(t)=e^{2\pi i t}-2^{1-s}e^{4\pi i t},
\]
a smooth zero-mean period-$1$ function.  For $0<\mathrm{Re}\,s<1$ and $k>0$ one has
\[
\int_0^\infty e^{ikt}\,t^{-s}\,dt=\Gamma(1-s)e^{i\pi(1-s)/2}k^{s-1},
\]
hence
\[
\int_0^\infty R_\lambda(t)\,t^{-s}\,dt
=\Gamma(1-s)e^{i\pi(1-s)/2}\big[(2\pi)^{s-1}-2^{1-s}(4\pi)^{s-1}\big]=0 .
\]
Define
\[
g_\lambda(x)=x^{(i\lambda-1)/2}\,\frac{i}{2}\int_0^x R_\lambda(t)\,t^{-(1+i\lambda)/2}\,dt ,
\]
which solves $(D-\lambda)g_\lambda=R_\lambda$ (integrating factor $x^{(1-i\lambda)/2}$).  At $0$, $R_\lambda(t)=R_\lambda(0)+O(t)$ with $R_\lambda(0)=1-2^{(1-i\lambda)/2}\neq0$ (since $|2^{(1-i\lambda)/2}|=\sqrt2\neq1$), the integral is $O(x^{1/2})$, and $g_\lambda$ is bounded.  At $\infty$ the vanishing moment gives
\[
g_\lambda(x)=-x^{(i\lambda-1)/2}\,\frac{i}{2}\int_x^\infty R_\lambda(t)\,t^{-(1+i\lambda)/2}\,dt ,
\]
and the tail is $O(x^{-1/2})$, so $|g_\lambda(x)|=O(x^{-1})$.  Hence $g_\lambda\in L^2(\R_+)$, $g_\lambda\not\equiv0$, and $g_\lambda\in\ker(T^*-\lambda)$ by Lemma~\ref{lem:S-defcriterion}.  The same factor $2^{1-s}$ works for every harmonic $m\ge1$: set
$R_\lambda^{(m)}(t)=e^{2\pi i m t}-2^{1-s}e^{4\pi i m t}$ and construct
$g_\lambda^{(m)}$ as above.  In any finite relation among the $R_\lambda^{(m)}$,
the largest frequency $2m$ forces the coefficient with largest $m$ to vanish;
descending induction proves independence.  Applying $D-\lambda$ then proves that
the corresponding $g_\lambda^{(m)}$ are independent.
\end{proof}

\section{The BBM eigenstates are absent from the completion}
\label{sec:S-eigenstates}

The BBM eigenfunctions are $\psi_z(x)=-\zeta(z,x+1)$ on $\R_+$, with eigenvalue $i(2z-1)$; the BBM quantization condition $\psi_z(0)=0$ is $\zeta(z)=0$~\cite{bbm2017}.

\begin{theorem}
\label{thm:S-eigenstates}
For $\mathrm{Re}\,z=\tfrac12$, $\psi_z$ does not define an element of $\widetilde{\mathcal H}_{\eta_0}$.
\end{theorem}

\begin{proof}
An element of $\widetilde{\mathcal H}_{\eta_0}$ is represented by an $\eta_0$-Cauchy sequence in $\Dom$, equivalently by its image under $\Delt$ in $L^2(\R_+)$; a smooth function $\psi$ defines such an element only if $\Delta\psi\in L^2(\R_+)$.  The Hurwitz identity $\zeta(z,x)-\zeta(z,x+1)=x^{-z}$ gives, for $x>1$,
\[
(\Delta\psi_z)(x)=\psi_z(x)-\psi_z(x-1)=-\zeta(z,x+1)+\zeta(z,x)=x^{-z} .
\]
For $\mathrm{Re}\,z=\tfrac12$ one has $|x^{-z}|=x^{-1/2}$ and $\int_1^\infty x^{-1}dx=\infty$.  Hence $\Delta\psi_z\notin L^2(\R_+)$.
\end{proof}

\begin{corollary}[Realization-independent obstruction]
\label{cor:S-kill}
No self-adjoint realization of the BBM formal expression in $\widetilde{\mathcal H}_{\eta_0}$ has the Riemann-zero ordinates with the BBM functions $\psi_z$ as eigenvectors: these functions are not vectors of the completed space, and the BBM boundary condition $\psi_z(0)=0$ is imposed on functions outside it.  Thus every self-adjoint extension in this completion fails to realize the original BBM candidate-eigenfunction/boundary-condition mechanism, even though some extensions may have unrelated point spectrum.
\end{corollary}

\section{Scope of the de Branges remark}

The theorems above concern the completion selected from the standard half-line $L^2$
core by $\Delta^\dagger\Delta$, and the no-go theorem for bounded sandwiches
$\Delta^\dagger h(D)\Delta$.  They neither exclude nor construct non-$L^2$ or
rigged-Hilbert-space formulations, unbounded metrics, distributional state spaces,
or genuinely new operator domains or eigenfunction ansatzes.  For the free
transported realization in Theorem~\ref{thm:S-main}, the $\Delta$ factors are absorbed
by $\Delt$; a further de Branges construction would then be a separate problem.
Other Hilbert--P\'olya Hamiltonians based on different operators and similarities,
such as Ref.~\cite{yakaboylu2024}, are likewise outside the present result.

For the natural candidate
\[
E(z)=\xi\left(\frac12+iz\right),
\]
one has
\[
E^\#(z)=\overline{E(\bar z)}=\xi\left(\frac12-iz\right)
=\xi\left(1-\left(\frac12-iz\right)\right)
=\xi\left(\frac12+iz\right)=E(z),
\]
using the functional equation $\xi(s)=\xi(1-s)$.  Hence the strict Hermite--Biehler inequality $|E(z)|>|E^\#(z)|$ for $\mathrm{Im}\,z>0$ fails.  This calculation rules out only this immediate choice of $E$; it is not a classification of de Branges realizations.

Any proposal along one of these routes would be a new construction rather than a
self-adjoint-extension rescue of the original BBM mechanism.  It would have to define
its topology, state space, and operator domain and establish self-adjointness and the
claimed spectrum independently.